\documentclass[useAMS,usenatbib]{mn2e}

\title[U Gem]{The masses, radii and luminosities of the components of 
U Geminorum}
\author[T. Naylor et al]{T. Naylor$^{1}$ A. Allan$^{1}$ and K.S.
Long$^{2}$\\
$^{1}$School of Physics, University of Exeter, Stocker Road, Exeter
EX4 4QL\\
$^{2}$Space Telescope Science Institute, 3700 San Martin Drive,
Baltimore, MD 21218, USA}
\begin{document}

\date{}

\pagerange{\pageref{firstpage}--\pageref{lastpage}} \pubyear{2005}

\maketitle

\label{firstpage}

\begin{abstract}
We present a phase-resolved spectroscopic study of the secondary star
in the cataclysmic variable U Gem.
We use our data to measure the radial velocity semi-amplitude,
systemic velocity and rotational velocity of the secondary star.
Combining this with literature data allows us to determine
masses and radii for both the secondary star and white dwarf which are
independent of any assumptions about their structure.
We use these to compare their properties with those of field stars and find
that both components follow field mass-radius relationships.
The secondary star has the mass, radius, luminosity and photometric 
temperature of an M2 star, but a spectroscopic temperature of M4.
The latter may well be due to a high metallicity.
There is a
troubling inconsistency between the radius of the white dwarf inferred from its
gravitational redshift and inclination and that inferred
from its temperature, flux, and astrometric distance.

We find that there are two fundamental limits to the accuracy of the 
parameters we can derive.
First the radial velocity curve of the secondary star deviates from a 
sinusoid, in part because of its asphericity (which can be modelled)
and in part because the line flux is not evenly distributed over its
surface.
Second we cannot be certain which spectral type is the best match
for the lines of the secondary star, and the derived rotational
velocity is a function of the spectral type of the template star used.

\end{abstract}


\section{INTRODUCTION}

Precise radii and masses for the white dwarfs and late-type secondary
stars in cataclysmic variables (CVs) are critical tests of our understanding
of their evolution.
For the secondary stars there has been a long debate as to whether
they follow main-sequence mass-radius and mass-temperature
relationships.
These issues are reviewed in \cite{1998MNRAS.301..767S}, who concluded
that, although the CV data are consistent with the main sequence
relationships, there is such a large scatter in the CV properties that
the main-sequence relationships cannot be used to predict one quantity
from another.
Amongst the CVs with M-type secondaries, a large part of that scatter
was driven by just two systems, IP Peg and U Gem.
However, these are systems whose derived parameters should be amongst the
most precise since they are eclipsing systems.

In an earlier paper we re-visited IP Peg \citep{2000MNRAS.318....9B}
and brought its parameters into line with those of the other CVs of
similar orbital period, making U Gem a natural next target.
However, there was another, pressing reason for re-visiting the
parameters of this system, in that \cite{1999ApJ...511..916L} obtained
a radial velocity curve for the white dwarf.  
This is the first radial velocity curve of a white dwarf obtained for any
CV.  This not only allows an almost assumption-free mass ratio
$q(=M_2/M_1)$ to be derived, but also allows one to measure the
gravitational redshift to the white dwarf, which given the white dwarf
mass yields its radius.
In principle, therefore, one could obtain independent measurements of
both the mass and radius for each of the white dwarf and secondary star.
There were three obstacles to such a complete analysis.
\begin{itemize}
\item{} There was no measurement of the rotational velocity of the
secondary star ($v_2 {\rm sin} (i)$) from which its radius could be
derived assuming only that it is phase locked.
\item{} To measure the gravitational redshift of the white dwarf
  requires a measurement of the mean radial velocity of the red dwarf 
  ($\gamma_2$).  There were two measurements of this which conflicted.
\item{} The radial velocity curve of the secondary star showed an
  apparent eccentricity.  The best explanation for this would be that
  the line flux was not distributed evenly across the surface of the 
star, but without more data to examine the residuals in detail, any
  corrections that should be applied to account for this were
  uncertain.
\end{itemize}
We therefore re-observed U Gem to solve these problems.
In doing so we acquired a very high quality dataset, which illuminates
various problems in parameter determination.
This was part of a programme to obtain phase resolved red spectroscopy of
cataclysmic variables and low-mass X-ray binaries with precise
correction for the telluric absorption, so that one could study the
TiO bands \citep{2002ApJ...568L..45W, 2000MNRAS.318....9B,
  2000MNRAS.317..528W}.  
The observational techniques used in this paper should be seen as a
further improvement on the ones described in those papers.

This paper is set out as follows.
In Sections \ref{obs} and \ref{extr} we describe the data acquisition
and reduction.
The almost complete absence of interlocking assumptions then allows us to
lay out the determination of the dynamical parameters in a very formal
manner, discussing each parameter and its uncertainties in turn
(Section \ref{dynam}).
Our dataset is such that we discover that the uncertainties in both 
$K_2$ and $v_2 {\rm sin} (i)$ are dominated by systematics.
We then use these to obtain the physical parameters of interest in
Section \ref{phys}, before comparing them to field stars in our
discussion (Section \ref{dis}).
We finally draw our conclusions in Section \ref{conc}.

\section{OBSERVATIONS}
\label{obs}

Spectra were obtained on the nights beginning January 4, 5, 6 and 7 2001 
and January 17 2000 using
the 2.5m Isaac Newton Telescope with the 235mm camera of the Intermediate
Dispersion Spectrograph using a 1.5 arcsec slit.
The slit width corresponds to approximately two pixels at the
detector, and was a compromise.
On the one hand narrowing it further would risk undersampling in
wavelength space, which would compromise the precision with which we
could measure velocities.
Conversely, widening it would mean that velocity shifts would be
introduced if the star moved across the width of the slit.
The largest possible value of these shifts corresponds to the slit
width, or $\pm$15km s$^{-1}$.
Since the $\gamma$ velocity of the white dwarf is only known to this
precision, and in practice the seeing would ensure the shifts were smaller than
this, the slit width was a reasonable compromise.  
In fact, as we shall discuss later, we found that these shifts could be 
measured and corrected, considerably improving the quality of our data.

The slit was orientated such that both U Gem and the $V\simeq15$ star
at $\alpha=$07 55 07.08 $\delta=$+21 59 19.7 (J2000) fell in the
slit. 
For the majority of our spectra we used the R1200R grating which gave
a resolution of 1.55\AA\ and a coverage of 7480-8320\AA\ at
0.83\AA/pixel.
This allowed us to cover the TiO band and KI lines at 7700\AA\ and the 
NaI doublet at 8190\AA.
On the last night we used the R1800V grating to cover 7855-8335\AA\ 
at 0.48\AA/pixel with a resolution of 0.76\AA, covering just the
NaI doublet.
We used an exposure of 420s for all our U Gem spectra.
Since we required accurate radial velocities we took an arc spectrum after
every two spectra of U Gem.
We were also concerned to ensure that the signal-to-noise in our final
mean spectra (which is around 350 for the lower resolution data) was not 
dominated by fringing on the CCD, so
at the same time as we acquired an arc, we also obtained a tungsten lamp
flatfield.

Since the primary aim of our observing procedure was to ensure excellent
removal of the telluric absorption, we interspersed our observations of
U Gem with observations of the rapidly rotating A0 star SAO78799
This star is relatively featureless in the range of interest, and so
could be used to correct the telluric bands (see Section \ref{extr}).
After each observation of the A0 star we took an arc spectrum and a flatfield
before moving the telescope.

We also required spectra of late-type stars for cross correlation and to
determine an accurate spectral type for U Gem.
These were obtained using the same instrumental setups as above.
Again each observation was accompanied by a flatfield and arc, along with
a spectrum of a nearby A0 star, which also had its own flatfield and arc.

\section{SPECTRUM EXTRACTION}
\label{extr}

We treated all our observations of U Gem and the late-type stars in an
identical fashion.
Mean bias frames were subtracted from each image.
We then created flatfields by calculating a correction with respect to
a two dimensional polynomial fitted through
each flatfield frame, and these were used to correct the arcs and star 
images.
Spectra were then extracted from each stellar observation using an optimal
technique \citep[e.g.][]{1986PASP...98..609H}.
For each such spectrum an arc was extracted from the appropriate exposure,
from the same region of the detector as the star was extracted from.
It was used to derive a polynomial relationship between pixel number 
and wavelength, which was used to wavelength calibrate the stellar
spectra.
Each target spectrum was then divided by the appropriate observation of 
an A0 star to remove the atmospheric absorption bands.

In an initial reduction of the data, we followed the telluric
correction procedure developed in \cite{2000MNRAS.318....9B}, where no
interpolation of the telluric correction star onto the wavelength scale
of the target was attempted as this can tend to smooth the spectra.
Instead each target spectrum pixel is simply divided by the nearest
pixel in wavelength from the A0 star spectrum.

In some cases this produced a rather poor telluric correction, and
this was traced to different positions of the target star across the
width of the slit, leading to small differences in the wavelength
calibration.
We therefore cross correlated all our target spectra and telluric
correction spectra with one A-star spectrum, in the region
7624-7690\AA\ which is dominated by telluric bandheads.
We then applied the derived shifts to the wavelength values of each
data point in all our target and telluric
correction spectra.
Before dividing we interpolated
the telluric correction stars onto the same wavelength points as their 
respective targets using quadratic interpolation.
We also normalised the correction spectrum such that it was one in the 
middle of its wavelength
range to preserve the overall number of counts correctly for the target.
This entire procedure improved some of the corrections dramatically.
Equally importantly, though, it corrects for the drift of U Gem across
the width of the slit, a point we shall return to in Section
\ref{K_2}.
We could not carry out this procedure for the higher resolution data,
as there is not a region of such strong telluric absorption.

Each spectrum was corrected to the velocity of the solar system 
barycentre by allowing for the Earth's motion. 
Throughout the reduction process we propagated the uncertainties from
the optimal extraction.
This allowed us to identify four spectra as of significantly lower
signal-to-noise than the others (probably due to cloud), which were
removed from the subsequent analysis.
A further spectrum was removed due to a charged particle event in
the region used for the velocity correction procedure.
This left us with 118 low resolution and 59 high resolution spectra.
Figure \ref{spectra} shows our final summed spectra, shifted into the 
reference frame of the secondary star.

The 118 lower resolution spectra, and corresponding mean late-type
star spectra are available via the CDS, or from 
http://www.astro.ex.ac.uk/people/timn/UGem/description.html.

\section{EXTRACTING THE DYNAMICAL PARAMETERS}
\label{dynam}

There are three parameters that we were interested in extracting from
our spectra.
The observed radial velocity semi-amplitude of the secondary star, before
correction for inclination ($K_2$); the rotational velocity of
the secondary star, again before allowing for inclination ($v_2
{\rm sin}(i)$); and the mean velocity of the secondary star ($\gamma_2$).
In principle we have to measure all three parameters simultaneously.
For example, we need a broadened spectral type template to
cross-correlate against our individual U Gem spectra to obtain $K_2$,
but can only obtain that by first velocity shifting all
the individual spectra into the rest frame of the secondary star and
summing them.
Of course we cannot do this without first knowing $K_2$.
We therefore adopted an iterative procedure outlined in the next two
paragraphs, with each stage described in detail in the subsections below.

We began by analysing the lower resolution spectra.
An inital radial velocity curve was extracted
by cross correlating the NaI region of the U Gem spectra against a
spectral type template.
A fit to this gave an initial radial velocity semi-amplitude which we
used to remove the orbital motion of the
secondary star by shifting each spectrum in velocity.
However, it was clear that the radial velocity curve was significantly 
non-sinusoidal between phases 0.25 and 0.75 (see Section
\ref{deviations} and Figures \ref{radvel} and \ref{res}),
and so we restricted these sums to the data outside this phase range.
We then measured the velocity shift between the summed spectrum
and the spectral type template, and then shifted the standard,
and broadened it so its apparent rotational velocity matched that
of the secondary star. 
We next re-cross correlated each U Gem spectrum with the broadened
template to create a new radial velocity curve, from which we could
again extract radial velocity curve parameters.
We iterated around this procedure until the change in the velocity 
shift we applied to the spectral type template converged to within one
km s$^{-1}$.
Having obtained an initial radial velocity solution from the NaI region using
one spectral type template
we then used this as the starting parameters for using each spectral
type template for both the NaI and KI/TiO regions.

As there are a smaller number of high resolution spectra, and they
could not be corrected in velocity for the slit drift (see Section 
\ref{extr}) we have used them only to measure $v_2 {\rm sin} (i)$.
Thus we 
coadded them using the $K_2$ found from the lower
resolution data, and then fitted broadened templates, again using
spectra from only phases 0.75 to 0.25.

\begin{figure}
\vspace{70mm}
\includegraphics{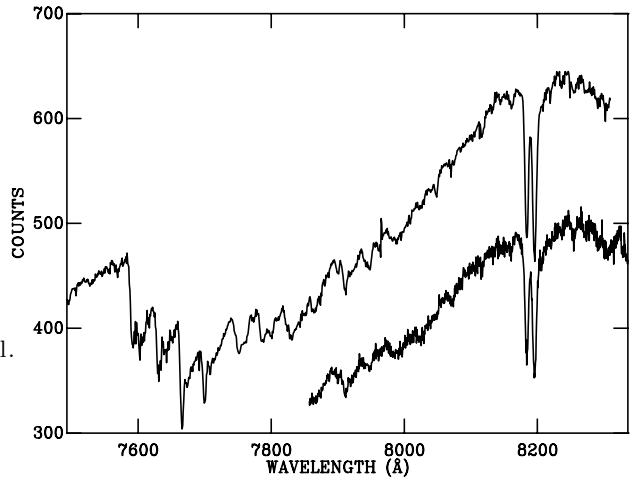}
\caption{
The mean of all spectra, summed into the rest frame of the secondary
star using the parameters derived from the NaI lines using GJ213
as the spectral type template.
The upper spectrum is the sum of the lower resolution data, the lower
spectrum the higher resolution data.
The higher resolution data have been multiplied by 1.8 (the difference
in pixel size in wavelength space).
}
\label{spectra}
\end{figure}

\begin{figure}
\vspace{70mm}
\includegraphics{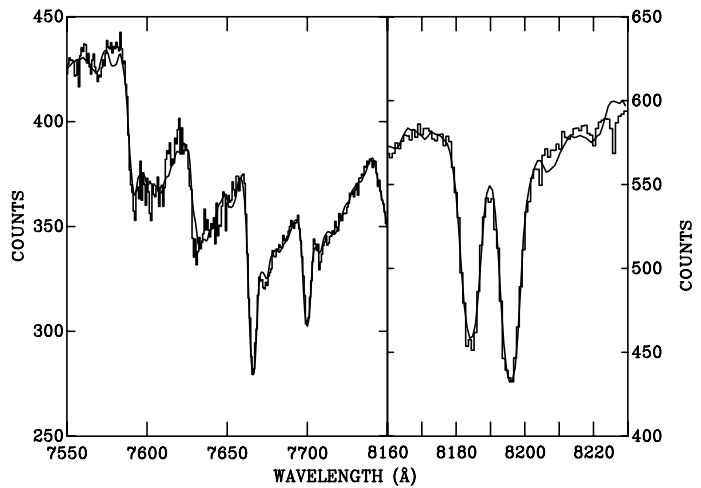}
\caption{
The mean of the low resolution spectra between phases 0.75 and 0.25
(histogram) with the best fit velocity broadened spectrum of GJ213
(line) overlaid.
The wavelength regions are those over which the fitting was performed.
}
\label{fit}
\end{figure}

Aside from the dynamical parameters, the final output from this
procedure is a spectrum in the rest frame of the secondary star.
We show these spectra for both the high and low resolution data in
Figure \ref{spectra}.
In addition, Figure \ref{fit} shows the comparison between a broadened
template star, and the sum of the spectra between phase 0.75 and 0.25
for the regions we used for the cross-correlation and $v_2 {\rm sin}
(i)$ fitting.

\begin{table}
\caption{Measured radial velocities from the NaI feature in the lower
  resolution spectra.  The columns give the barycentric Julian Date
  for the middle of the exposure, the calculated phase for the
  exposure, the image number, radial velocity and uncertainty.  The
  uncertainties are scaled to give a $\chi^2_\nu$ of one for a
  circular fit.  All exposures are of 420s duration.}
\begin{tabular}{ccccccccccccccccc}
\hline
BJD & Phase & Image & RV & Uncertainty \cr
    &       & Number & (km s$^{-1}$)   &(km s$^{-1}$)\cr
\hline
2\ 451\ 561.43994 & 0.5544 & 206090 & -122 & 34\cr
2\ 451\ 561.45081 & 0.6158 & 206094 & -219 & 32\cr
2\ 451\ 561.45612 & 0.6458 & 206095 & -247 & 29\cr
2\ 451\ 914.41589 & 0.8262 & 243111 & -253 & 11\cr
2\ 451\ 914.42090 & 0.8545 & 243112 & -224 & 10\cr
2\ 451\ 914.42700 & 0.8890 & 243115 & -185 & 11\cr
2\ 451\ 914.43201 & 0.9173 & 243116 & -141 & 11\cr
2\ 451\ 914.44177 & 0.9725 & 243123 &  -49 & 10\cr
2\ 451\ 914.44672 & 0.0005 & 243124 &    4 & 11\cr
2\ 451\ 914.45264 & 0.0339 & 243127 &   63 & 10\cr
2\ 451\ 914.45758 & 0.0619 & 243128 &  121 &  9\cr
2\ 451\ 914.46594 & 0.1091 & 243135 &  193 &  9\cr
2\ 451\ 914.47095 & 0.1374 & 243136 &  239 &  9\cr
2\ 451\ 914.47687 & 0.1709 & 243139 &  262 &  9\cr
2\ 451\ 914.48181 & 0.1988 & 243140 &  279 &  8\cr
2\ 451\ 914.49042 & 0.2475 & 243146 &  299 &  9\cr
2\ 451\ 914.49542 & 0.2758 & 243147 &  293 &  9\cr
2\ 451\ 914.50122 & 0.3085 & 243150 &  280 &  8\cr
2\ 451\ 914.50616 & 0.3365 & 243151 &  268 &  8\cr
2\ 451\ 914.51453 & 0.3838 & 243157 &  210 & 11\cr
2\ 451\ 914.51947 & 0.4117 & 243158 &  160 & 10\cr
2\ 451\ 914.52527 & 0.4445 & 243161 &  106 & 10\cr
2\ 451\ 914.53027 & 0.4728 & 243162 &   53 & 12\cr
2\ 451\ 914.53937 & 0.5242 & 243168 &  -65 &  9\cr
2\ 451\ 914.54437 & 0.5525 & 243169 & -125 &  8\cr
2\ 451\ 914.55017 & 0.5853 & 243172 & -175 &  9\cr
2\ 451\ 914.55518 & 0.6135 & 243173 & -218 &  8\cr
2\ 451\ 914.56470 & 0.6674 & 243180 & -266 & 10\cr
2\ 451\ 914.56964 & 0.6953 & 243181 & -293 &  8\cr
2\ 451\ 914.57550 & 0.7284 & 243184 & -302 &  8\cr
2\ 451\ 914.58044 & 0.7564 & 243185 & -282 &  8\cr
2\ 451\ 914.58881 & 0.8036 & 243191 & -283 &  8\cr
2\ 451\ 914.59375 & 0.8316 & 243192 & -251 &  8\cr
2\ 451\ 914.59961 & 0.8647 & 243195 & -217 &  8\cr
2\ 451\ 914.60455 & 0.8927 & 243196 & -188 &  8\cr
2\ 451\ 914.61292 & 0.9399 & 243202 & -118 &  8\cr
2\ 451\ 914.61786 & 0.9679 & 243203 &  -73 &  8\cr
2\ 451\ 914.62372 & 0.0010 & 243206 &    3 &  9\cr
2\ 451\ 914.62866 & 0.0289 & 243207 &   60 & 12\cr
2\ 451\ 914.63660 & 0.0738 & 243213 &  131 & 11\cr
2\ 451\ 914.64160 & 0.1021 & 243214 &  181 & 11\cr
2\ 451\ 914.64758 & 0.1359 & 243217 &  224 & 13\cr
2\ 451\ 914.65253 & 0.1638 & 243218 &  254 & 10\cr
2\ 451\ 914.65753 & 0.1921 & 243219 &  274 &  9\cr
2\ 451\ 914.66583 & 0.2391 & 243225 &  282 &  9\cr
2\ 451\ 914.67084 & 0.2673 & 243226 &  285 & 11\cr
2\ 451\ 914.67657 & 0.2998 & 243229 &  287 &  9\cr
2\ 451\ 914.68158 & 0.3281 & 243230 &  269 &  8\cr
2\ 451\ 914.68988 & 0.3750 & 243236 &  224 &  8\cr
2\ 451\ 914.69482 & 0.4029 & 243237 &  182 &  8\cr
2\ 451\ 914.70062 & 0.4357 & 243240 &  128 &  8\cr
2\ 451\ 914.70557 & 0.4637 & 243241 &   83 &  8\cr
2\ 451\ 914.77521 & 0.8573 & 243272 & -227 & 10\cr
2\ 451\ 914.78015 & 0.8853 & 243273 & -192 & 11\cr
2\ 451\ 914.79144 & 0.9491 & 243280 &  -91 & 12\cr
2\ 451\ 914.79639 & 0.9770 & 243281 &  -36 & 12\cr
2\ 451\ 915.44385 & 0.6369 & 243399 & -233 &  9\cr
2\ 451\ 915.44879 & 0.6649 & 243400 & -255 &  9\cr
2\ 451\ 915.45428 & 0.6959 & 243403 & -292 &  9\cr
\hline
\end{tabular}
\label{vel_data}
\end{table}

\begin{tabular}{ccccccccccccccccc}
\hline
BJD & Phase & Image & RV & Uncertainty \cr
    &       & Number & (km s$^{-1}$)   &(km s$^{-1}$)\cr
\hline
2\ 451\ 915.45923 & 0.7239 & 243404 & -292 &  8\cr
2\ 451\ 915.46637 & 0.7643 & 243408 & -289 &  8\cr
2\ 451\ 915.47131 & 0.7922 & 243409 & -287 &  8\cr
2\ 451\ 915.47687 & 0.8236 & 243412 & -262 &  8\cr
2\ 451\ 915.48181 & 0.8515 & 243413 & -235 &  8\cr
2\ 451\ 915.48926 & 0.8936 & 243417 & -184 &  8\cr
2\ 451\ 915.49420 & 0.9216 & 243418 & -137 &  8\cr
2\ 451\ 915.49976 & 0.9530 & 243421 &  -97 &  9\cr
2\ 451\ 915.50470 & 0.9809 & 243422 &  -33 & 11\cr
2\ 451\ 915.51202 & 0.0223 & 243426 &   43 & 18\cr
2\ 451\ 915.51697 & 0.0503 & 243427 &   63 & 39\cr
2\ 451\ 915.52252 & 0.0817 & 243430 &  147 & 19\cr
2\ 451\ 915.52747 & 0.1096 & 243431 &  208 & 20\cr
2\ 451\ 915.53467 & 0.1503 & 243435 &  251 & 21\cr
2\ 451\ 915.53961 & 0.1783 & 243436 &  284 & 23\cr
2\ 451\ 915.55011 & 0.2376 & 243440 &  311 & 17\cr
2\ 451\ 915.56897 & 0.3442 & 243445 &  253 & 14\cr
2\ 451\ 915.57391 & 0.3722 & 243446 &  226 & 12\cr
2\ 451\ 915.57941 & 0.4032 & 243449 &  192 & 13\cr
2\ 451\ 915.58429 & 0.4308 & 243450 &  135 & 12\cr
2\ 451\ 916.43011 & 0.2120 & 243541 &  287 &  9\cr
2\ 451\ 916.43506 & 0.2400 & 243542 &  293 &  9\cr
2\ 451\ 916.44061 & 0.2714 & 243545 &  290 &  9\cr
2\ 451\ 916.44550 & 0.2990 & 243546 &  291 &  9\cr
2\ 451\ 916.45367 & 0.3452 & 243550 &  248 &  9\cr
2\ 451\ 916.45862 & 0.3732 & 243551 &  220 &  9\cr
2\ 451\ 916.46417 & 0.4046 & 243554 &  201 & 24\cr
2\ 451\ 916.50348 & 0.6267 & 243559 & -231 &  9\cr
2\ 451\ 916.50842 & 0.6547 & 243560 & -256 &  8\cr
2\ 451\ 916.51398 & 0.6861 & 243563 & -279 &  8\cr
2\ 451\ 916.51892 & 0.7140 & 243564 & -297 &  8\cr
2\ 451\ 916.52625 & 0.7554 & 243568 & -291 & 11\cr
2\ 451\ 916.53119 & 0.7834 & 243569 & -286 &  9\cr
2\ 451\ 916.53674 & 0.8148 & 243572 & -271 &  8\cr
2\ 451\ 916.54169 & 0.8427 & 243573 & -247 &  8\cr
2\ 451\ 916.54938 & 0.8862 & 243577 & -200 &  9\cr
2\ 451\ 916.55432 & 0.9141 & 243578 & -159 &  9\cr
2\ 451\ 916.55981 & 0.9452 & 243581 & -101 & 10\cr
2\ 451\ 916.56482 & 0.9735 & 243582 &  -55 &  9\cr
2\ 451\ 916.57239 & 0.0163 & 243586 &   20 & 10\cr
2\ 451\ 916.57733 & 0.0442 & 243587 &   63 &  9\cr
2\ 451\ 916.58282 & 0.0753 & 243590 &  141 &  8\cr
2\ 451\ 916.58777 & 0.1032 & 243591 &  174 &  8\cr
2\ 451\ 916.59698 & 0.1553 & 243595 &  242 &  8\cr
2\ 451\ 916.60193 & 0.1832 & 243596 &  277 &  8\cr
2\ 451\ 916.60748 & 0.2146 & 243599 &  289 &  8\cr
2\ 451\ 916.61243 & 0.2426 & 243600 &  286 &  8\cr
2\ 451\ 916.63782 & 0.3861 & 243632 &  209 &  8\cr
2\ 451\ 916.64276 & 0.4141 & 243633 &  162 &  9\cr
2\ 451\ 916.64838 & 0.4458 & 243636 &  105 &  9\cr
2\ 451\ 916.65338 & 0.4741 & 243637 &   53 &  9\cr
2\ 451\ 916.66052 & 0.5145 & 243641 &  -28 & 10\cr
2\ 451\ 916.66553 & 0.5428 & 243642 &  -96 & 14\cr
2\ 451\ 916.67114 & 0.5745 & 243645 & -152 & 12\cr
2\ 451\ 916.67615 & 0.6028 & 243646 & -193 & 16\cr
2\ 451\ 916.68915 & 0.6763 & 243651 & -267 & 11\cr
2\ 451\ 916.69476 & 0.7080 & 243654 & -275 & 11\cr
2\ 451\ 916.69977 & 0.7363 & 243655 & -306 & 20\cr
2\ 451\ 916.70721 & 0.7784 & 243659 & -269 & 25\cr
\hline
\end{tabular}

\subsection{$\bf K_2$}
\label{K_2}

\subsubsection{The radial velocity curve}

\begin{figure}
\vspace{70mm}
\includegraphics{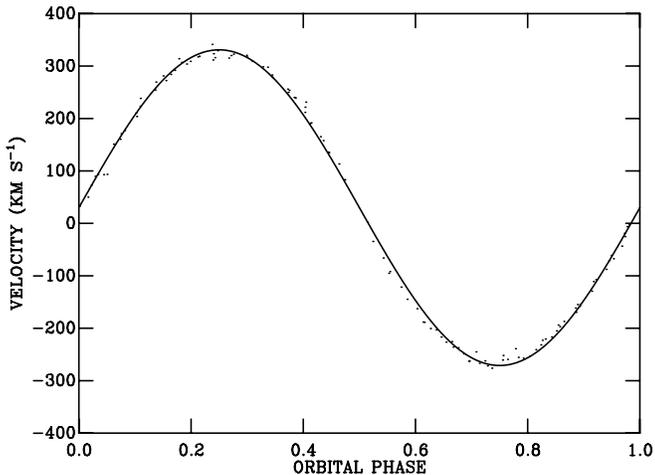}
\caption{
The radial velocity curve extracted from the low resolution data by
cross-correlation of the NaI lines with the M4.0 star GJ213,
corrected to the solar system barycentre.
The curve is the best (circular) fit to the data.
}
\label{radvel}
\end{figure}

\begin{figure}
\vspace{70mm}
\includegraphics{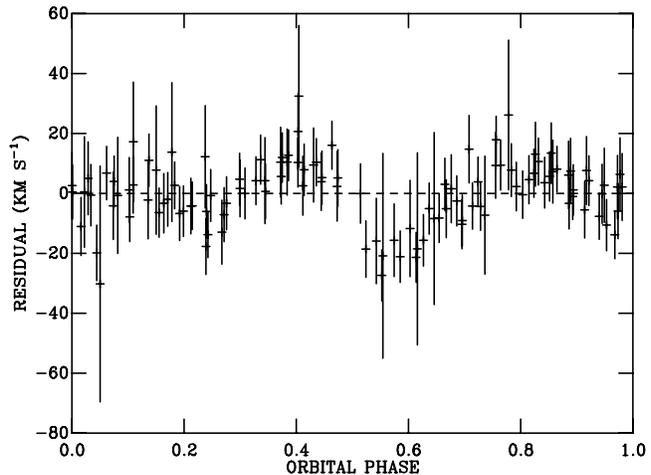}
\caption{
The residuals about the fit shown in Figure \ref{radvel}.
The error bars are the relative uncertainties scaled to give a
$\chi^2_{\nu}$ of one.
}
\label{res}
\end{figure}

We cross-correlated the KI/TiO (7550-7750\AA) and NaI (8160-8230\AA)
regions of each
spectrum with that of each spectral type template.
We chose the upper limit of the TiO region to avoid the feature at
7773\AA, which is probably OI \citep{1988MNRAS.233..451F}.
We then fitted a sine curve to the resulting radial velocity curves,
with the period fixed at that of \cite{1990ApJ...364..637M}. 
Each velocity was weighted according to the calculated
signal-to-noise in the spectrum from which it was derived, using the
region 7800-8000\AA\ for the low resolution
data and 8140-8250\AA\ for the high resolution data.
Since we only have relative weights for the data points, we cannot
assign absolute values of $\chi^2$, but the relative change in
$\chi^2$ between the different fits is meaningful.
We found that all the NaI fits have $\chi^2$ values within $\pm$2 percent
of each other, but the KI/TiO region gives values 20 to 80 percent larger.
We therefore decided that NaI provides a better solution, and 
in Figure \ref{radvel} we show the radial velocity curve and the best fit and in
Figure \ref{res} the residuals about that fit.
Table \ref{vel_data} gives the individual radial velocity points.

We note in passing that the RMS of the fits fell considerably between
the initial and final radial velocity curves.
This is presumably because the broadened templates match the data
better, and so yield better cross-correlation functions.
This leads us to believe that the reason we obtained better
results in \cite{2000MNRAS.318....9B} from cross-correlating with a
summed spectrum of IP Peg rather than the template stars was that the
latter were not broadened.

\subsubsection{The nature of the deviations}
\label{deviations}

It is immediately obvious from Figures \ref{radvel} and \ref{res} that 
there are systematic deviations from the best fit. 
There are two obvious hypotheses which might explain these deviations.
The first is a region of reduced NaI absorption flux around the inner 
Lagrangian point.
There is observational backing for such an hypothesis, since
surface mapping of the secondary star using the radial
velocity data available prior to the current work, maps the observed
apparent eccentricity into a region of reduced NaI absorption
\citep{1992MNRAS.257..476D}.
Such a region would lead to a lack of
NaI absorption on the hemisphere of the secondary star rotating towards
the observer at phases around 0.4, and away from the observer at 0.6.
Since removing flux from the blueward wing of the line will result in
an apparent redshift, this would, apparently, explain the residuals.
There are a range of possible physical causes for such a region, but
it is unclear whether the region needs to be hotter or cooler than the
immaculate photosphere to cause a deficit in NaI absorption.
Heating will cause the overall continuum flux to rise, but will also 
decrease the equivalent width of the NaI line.
Which effect will dominate the change in flux is unclear, though the
irradiation calculations of \cite{1993MNRAS.264..641B} do yield the 
required decrease in NaI flux.
Cool spots on the other hand may be caused by by magnetic activity
\citep[see the observations of][and references
therein]{2002ApJ...568L..45W}, or by a combination of limb and gravity
darkening.
Again the sign of the effect on the NaI flux is unclear as in the
former case there is is also a gravity effect, and in the latter the
line limb and gravity darkening co-efficients are essentially unknown.
Nevertheless, it appears that a spot resulting in a decrease in the
NaI (and KI) flux at the stellar surface could explain the
observations.

The second obvious candidate is absorption of light from the secondary
star as it passes through the accretion disc.
\cite{2001MNRAS.327..475L} have demonstrated the importance of such
``mirror eclipses'' in the IR, and finding their counterpart in
optical data would clearly be of importance.
A mirror eclipse would create further absorption in the spectrum.
At phase 0.4, the star would be moving behind material moving away
from the observer, and at phase 0.6 material moving towards them.
This excess absorption would again create the observed effect.

\begin{figure}
\vspace{70mm}
\includegraphics{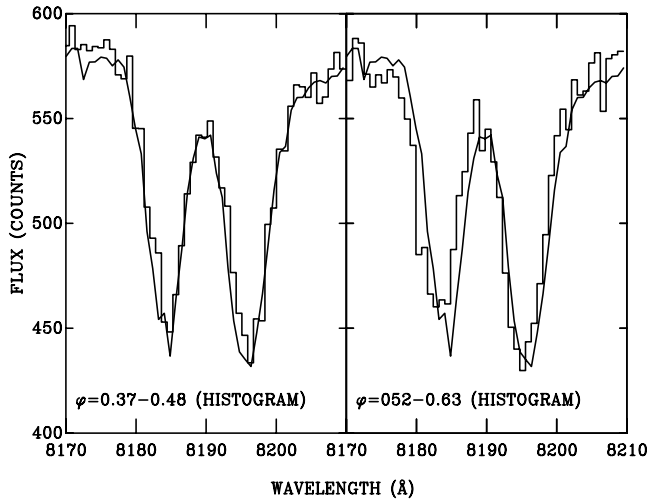}
\caption{
The mean of the spectra between phase 0.75 and 0.25 (solid line)
compared with the mean of spectra in the phase range 
0.37 to 0.48 (left) and 0.52 to 0.63 (right).
}
\label{both}
\end{figure}

There is a very obvious way to choose between these models, by
comparing the spectra at phases 0.4 and 0.6 with our summed spectrum
taken from phases 0.75 to 0.25.
The dark spot model predicts, at phase 0.4, that there will be a lack
of flux on the blue wings of the line, whilst a mirror eclipse would
predict extra absorption on the red wing.
Figure \ref{both} shows these spectra, scaled so that the continuum
values are similar to the phase 0.75 to 0.25 sum, which has been
overlaid on the same plots.
It is reassuring that with such a simple scaling the depths of the NaI
lines match so well; it implies that the majority of the variability
and of the flux originates from the secondary star.
If this was not the case the equivalent width of the NaI lines would
not be so constant.

The case at phase 0.4 seems clear-cut.
The red wings of the lines are co-incident, but the blue wing of
the phase 0.4 spectrum is consistently redder by between a half and 
one pixel (15-30 km s$^{-1}$).
This lack of blue absorption at this phase supports the idea of a spot.
At phase 0.6 the situation is more ambiguous, with the whole line
being shifted bluewards by between a half and one and a half pixels
(15-45 km s$^{-1}$).

\begin{figure}
\vspace{70mm}
\includegraphics{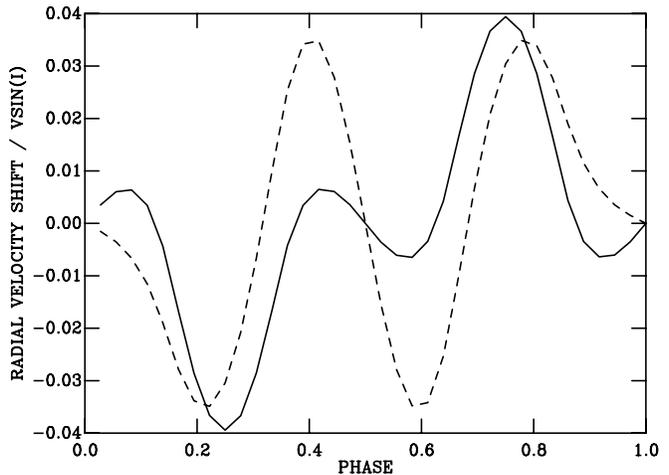}
\caption{
The radial velocity shifts expected from two different models of the
secondary star.
The solid line is for a uniform distribution of line flux over a
Roche-lobe filling secondary star for a $q$ of 1/3.
The dashed line shows the result of adding gravity darkening. 
}
\label{shift_model}
\end{figure}

To assess the credibility of the spot model further, we modelled the
radial velocity curve of a Roche-lobe filling star using a modified
version of the code first presented in \cite{1993MNRAS.265..655S}.
We assigned a radial velocity to each point on the grid which covers
the surface of the secondary star and used the fluxes from each
point to produce a line profile at that phase.
As is usual in modelling rotational profiles, we normalised the
velocity scale for the profile in terms of $v_2{\rm sin}(i)$.
Throughout the modelling we used a mass ratio $q=M_2/M_1=$1/3, and inclination 
of 67 degrees.
We calculated the radial velocity deviations the model would produce 
from a circular orbit as the first moment of the profile.
As a baseline, we first calculated a model with no limb or gravity
darkening.
Such a model is equivalent to assuming that the NaI flux is uniform
across the surface of the star, and does not depend on the angle an
element is viewed at.
Figure \ref{shift_model} shows the result, with deviations of the
correct order (0.04$v_2{\rm sin}(i)\simeq$12km s$^{-1}$), but at the
wrong phases (0.25 and 0.75).
These deviations are due to the asphericity of the secondary star.
The inner Lagrangian point moves faster than the mean surface of the
secondary, and so when on the limb of the secondary moving towards the
observer (as at phase 0.25) it produces a small blue shift with respect
to the expected radial velocity curve.
This effect can be observed in our residuals in Figure \ref{res} but
is not the dominating ``problem''.

We next placed a spot on the inner Lagrangian point.
Whilst we could have achieved this by an arbitrary dimming of
the flux, we chose instead to include gravity darkening in our model,
by using a ``standard'' gravity darkening co-efficient for a
convective envelope of 0.08 \citep{1967ZA.....65...89L}.
We did this not with the idea of establishing that gravity darkening
is the cause of the dark spot, but simply to establish whether it has 
the correct order-of-magnitude.
The result of this is again shown in Figure \ref{shift_model}, where
it can be seen that it results in two more peaks in the radial
velocity curve residual, of roughly the same size as those due to the
asphericity of the secondary star, but now at phase 0.4 and 0.6.

\begin{figure}
\vspace{70mm}
\includegraphics{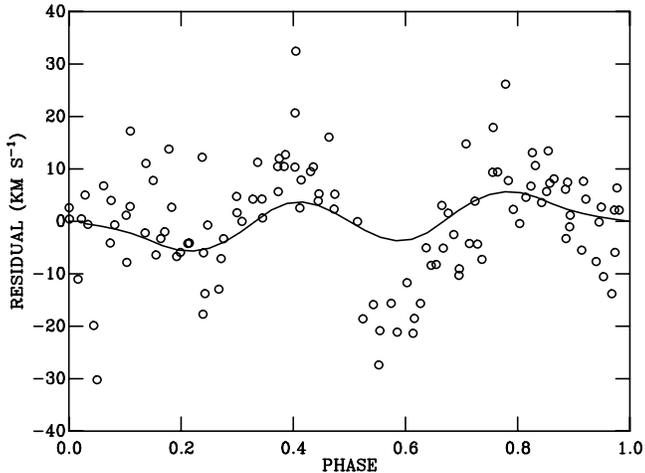}
\caption{
The solid line is the model for the residuals which includes both the
Roche geometry and gravity darkening (the dashed line of Figure
\ref{shift_model}).
We have subtracted from this model the best fitting sine wave with the
phase fixed, and then scaled it using a $v_2$sin$(i)$ of 115 km s$^{-1}$.
The circles are the residuals presented in Figure \ref{both}.
}
\label{it_works}
\end{figure}

Before comparing this model with our residuals we subtracted from it
the best fitting sinusoid with a fixed phase zero to simulate the
effect of our fitting process.  This has an amplitude of 0.01$v_2{\rm
sin}(i)$.
The result is shown in Figure \ref{it_works}, where we have assumed a
$v_2{\rm sin}(i)$ of 115 km s$^{-1}$ (see Section \ref{rot}).
The deviations due to the asphericity of the secondary star are well 
modelled, but it is clear we need a somewhat darker spot to reproduce
the data.
Furthermore, the dark spot deviations are obviously asymmetric, which cannot be
explained by this simple model.
However, further modelling lies beyond the scope of this work, our aim
is simply to establish how far these deviations from a perfect sinusoid
affect our derived parameters.

\subsubsection{Deciding a value and an uncertainty}
\label{k2_uncer}

The important result from Section \ref{deviations} is that the result
of fitting a sinusoid to  
the radial velocity deviations due to both the asphericity of the
secondary star, and the inner Lagrangian dark spot is about 0.01$v_2{\rm
sin}(i)$, or about 1.2 km s$^{-1}$.
Clearly this is not the entire story, it looks as though the dark spot
is darker than our predictions, and the asymmetry remains
unexplained.
We therefore apply our predicted correction, since the effects
modelled in it should be present, but we should also be aware that
there are probably further corrections of a similar order.
If we simply take a mean of the NaI values in Table \ref{results} and
make this correction our final answer is 300 km s$^{-1}$.
Deciding an uncertainty for this value is not trivial.
We can derive a statistical uncertainty for a the sine fit 
by scaling the relative error bars to yield a $\chi_{\nu}^2$ of 1,
which yields around 1 km s$^{-1}$.
Of course, if we included the effects of asphericity and the dark spot
in our model the size of error bar required to reach a $\chi_{\nu}^2$
of one would decline, along with our final uncertainty estimate.
Conversely, it is clear that our symmetric model will never fit both
the deviations at phase 0.4 and 0.6 to better than 5km s$^{-1}$.
Even if we used that as an uncertainty, it would never dominate the
uncertainties in our final parameters for U Gem.
Therefore we will ignore the uncertainty in $K_2$
throughout this paper, and leave any future users of this measurement 
to use the above discussion to choose an appropriate uncertainty for their use.

\subsubsection{Comparison with previous values}

We can now, finally, understand the apparent eccentricities found by
\cite{1990MNRAS.246..637F} and \cite{1981ApJ...246..215W} of 0.027
and 0.086 respectively.
If we fit our own data with an elliptical orbit we also obtain an
eccentricity of about 0.024, with a significant decrease in the
residuals caused by the inner Lagrangian dark spot.
So, it seems clear that the measured eccentricities in the orbit of U
Gem are caused by these effects, but that the correction for this,
derived in the previous section seems to be small, $\sim$1 km s$^{-1}$.
Therefore, to compare our value for the radial velocity
semi-amplitude of the secondary star with previous work, it seems best 
to use the raw $K_2$, i.e. without the corrections
suggested either by ourselves or \cite{1990MNRAS.246..637F}.
Our value of 302km s$^{-1}$ should therefore be compared with
$309\pm3$km s$^{-1}$ \citep{1990MNRAS.246..637F} and
$283\pm15$ \citep{1981ApJ...246..215W}.
Given the uncertainty we have about our errors, these values are in agreement.

\subsection{Spectroscopic Phase Zero}

Although this is not important for our work here, we can also use our
radial velocity curve to derive a
measurement of spectroscopic phase zero.
We obtain a time of TDB 2 451 915.8618
or HJD 2 451 915.8611.  The uncertainty is about 0.0002 day.
This lies about 0.0006 day or about 1.5$\sigma$ away from the
extrapolation of the \cite{1990ApJ...364..637M} ephemeris, after
taking into account the uncertainties in the ephemeris itself.

\subsection{V$\bf _2$sin(i)}
\label{rot}

\subsubsection{Method}

Examination of Figure \ref{res} shows that there are significant
deviations from the radial velocity curve between phases 0.25 and
0.75, as one might expect as a result of an inner Lagrangian dark
spot.
This implies that the dark spot significantly distorts the line
profile, and so if we are to determine the rotational velocity of the
secondary star it is sensible to omit these phases.
Therefore we co-added the spectra between phases 0.75 and 0.25 using
the derived semi-amplitude, and then fitted them in the same
wavelength regions used to derive $K_2$, to a broadened 
version of the spectral type templates, along with a smoothly varying 
continuum.
We broadened the spectral type template using the analytical function
appropriate for a spherical star.

Each of our spectra is a time average, during which the secondary star
will move along its orbit, adding a further broadening to the line
profiles.
To check that we had succeeded in making our observations short enough
to ensure this effect was negligible we calculated the maximum
extra broadening any spectrum should have (30 km s$^{-1}$).
We then pre-broadened the spectrum of the spectral-type standard by
this amount, and repeated our fitting procedure.
This changed the derived $v_2 {\rm sin}(i)$ by $<$ 4 km s$^{-1}$, and so
the effects of velocity smearing can be neglected.

As discussed in Section \ref{deviations}, the non-sphericity of the 
secondary star has a significant effect on the radial velocity curve.
\cite{1998MNRAS.298..153S} shows that we might also expect it to
affect the line profile broadening, making it significantly different
from the spherical star broadening assumed in our model.
To test this we used our Roche model to create a series of line
profiles from phase 0.25 to 0.75, and averaged them.
We then fitted this profile with our spherical model, and found that
the difference in derived $v_2{\rm sin}(i)$ was less than 2 percent.
This is far smaller than other uncertainties in our analysis and so
can be safely ignored.

\subsubsection{Deciding a value and an uncertainty}

The statistical uncertainties in our values can be estimated by
creating a $\chi^2$ grid for the two parameters, which are the fraction
of the light from the secondary star and the rotational broadening.
At each point in the grid we calculated the value of $\chi^2$, and
then re-normalised the grid so that the best fitting model gave a
$\chi^2_\nu$ of one.
Our 1$\sigma$ uncertainty is then simply the range of values of $v_2 {\rm
  sin}(i)$ which lie within a $\chi^2$ (not $\chi^2_\nu$) of 2.3 of the
best fit.
This procedure is conservative in the sense that the calculated
uncertainty would be smaller if we did not rescale $\chi^2$.
This gives uncertainties around 10 km s$^{-1}$ for either of the two
regions used at low resolution, or the high resolution data.

In addition to the statistical uncertainty, we must also consider the
fact that there is a very obvious correlation between the values of
$v_2{\rm sin}(i)$ given in Table \ref{results} and the spectral type
of the template used for the cross-correlation.
Later spectral types gave smaller values of $v_2{\rm sin}(i)$,
presumably because the lines in these stars are closer to saturation,
and so need less broadening to match the width of the lines of U Gem.
As we shall discuss in Section \ref{sec_star_disc} we cannot use the spectral
type of the secondary star as a guide, and so we are forced to look at
the values of $\chi_{\nu}^2$ to decide which fits are the most
appropriate.
The best NaI fits cover the range 120-130km s$^{-1}$, whilst KI/TiO
cover 105-130 km s$^{-1}$.
We therefore adopt a value of 115$\pm$15km s$^{-1}$ for $v_2{\rm
  sin}(i)$, where the uncertainty is driven by not knowing the correct
template to use, as this is larger than the uncertainty derived from
the $\chi^2$ analysis.

\begin{table*}
\caption{Measured Parameters.  Columns 1-3 give the spectral type,
  name and assumed $\gamma$ velocity for each spectral type standard.
Columns 4-5 give the values of $K_2$ derived from the NaI and KI/TiO
features in the lower resolution spectra.
Columns 6-11 give the results for velocity broadening the spectral
type templates to match the mean U Gem spectrum.  
For each parameter we give the values derived from the NaI and
K{I}/TiO features in the lower resolution spectra, and from the NaI feature in
the higher resolution spectra.
Thus columns 6-8 give the best fitting values of $v_2{\rm sin}(i)$ and (in brackets) the associated 
$\chi_{\nu}^2$,
and columns 9-11
the implied fraction of the light originating from the secondary star.
Columns 12-14 give the derived values of $\gamma_2$ for U Gem, again
from the lower resolution NaI and KI/TiO range, and the higher resolution
NaI range.}
\begin{tabular}{llccccccccccccccc}
\hline
\multicolumn{3}{c}{Radial Velocity standard } & \multicolumn{2}{c}{$K_2$ } & \multicolumn{3}{c}{$v_2{\rm sin}(i)$}      & \multicolumn{3}{c}{\% from} & \multicolumn{3}{c}{$\gamma_2$ }\\
Type  & Name     & $\gamma$      & \multicolumn{2}{c}{(km s$^{-1}$)} & \multicolumn{3}{c}{(km s$^{-1}$)} & \multicolumn{3}{c}{secondary   }  &\multicolumn{3}{c}{(km s$^{-1}$)}\\
      &          & (km s$^{-1}$) &     NaI & K{I}/TiO                  & NaI & K{I}/TiO & Hi                 &  NaI & K{I}/TiO & Hi & NaI & K{I}/TiO & Hi \\
(1) & (2) & (3) & (4) & (5) & (6) & (7) & (8) & (9) & (10) & (11) & (12) & (13) & (14) \cr
\hline
M3       & GJ463 &  21  & 302  &  305  & 142 (3.1) &  143 (2.8) &           & 116 & 109 &     & 38 & 43 \cr       
M3       & GL109 &  30  & 301  &  303  & 139 (2.9) &  140 (2.4) &           & 104 &  84 &     & 30 & 34 \cr       
M4       & GL490B& -23  & 301  &  303  & 130 (2.3) &  116 (2.3) &           &  76 &  51 &     & 11 & 18 \cr       
M4       & GJ213 & 106  & 301  &  303  & 132 (2.7) &  129 (2.3) & 121 (1.0) &  98 &  64 &  77 & 18 & 30 & 30 \cr  
M5       & GL51  &  -7  & 302  &  300  & 119 (2.6) &  105 (2.2) & 118 (1.1) &  58 &  38 &  67 & 24 & 37 & 25 \cr  
M5.5     & GL65A &  22  & 301  &  296  & 118 (2.9) &   96 (2.7) & 106 (1.1) &  61 &  31 &  53 & 25 & 39 & 28 \cr  
M6       & GJ1111&   8  & 301  &  289  & 110 (3.1) &  128 (3.4) & 110 (1.1) &  50 &  26 &  44 & 43 & 66 & 48 \cr  
\hline
\multicolumn{7}{l}{Mean (GL109, GJ213, GL51, GL65A)} &&&&& 24 & 34 & 27 \\
\multicolumn{7}{l}{Standard error}                   &&&&&  5 &  4 &  3 \\
\hline
\end{tabular}
\label{results}
\end{table*}

\subsection{$\bf \gamma_2$}

\subsubsection{Method}

To obtain $\gamma_2$ we calculated the difference in velocity between
each spectral type template and the mean, velocity shifted U Gem
spectrum it produced, and then added the barycentric radial velocity
of the template obtained from the literature.
The obvious course is to take a simple mean of all the derived $\gamma$ 
velocities, but before doing so we will examine the values obtained
as a function of both lines measured, and spectral type standard used.

The most obviously discrepant values are those given by GJ1111.
As indicated by the values of $\chi_{\nu}^2$ in Table \ref{results},
the broadened template for this star is a very bad fit to the data.
The next most discrepant values are those for GL490B.
Here the discrepancy is almost certainly due to the $\gamma$ velocity
assumed for GL490B.
Most of the template star velocities are
from \cite{2002AJ....123.3356G}, and have a accuracy of better than
1.5 km s$^{-1}$, but two objects 
(GJ463 and GL490B) are from \cite{1995AJ....110.1838R}.
Although \cite{1995AJ....110.1838R} estimate the accuracy of their
velocity measurements at 10km s$^{-1}$, we calculated the differences
between a given star's measured velocity in \cite{1995AJ....110.1838R}
and that in \cite{1987PASP...99..490M}.
The velocities in the latter have uncertainties of less than 1km s$^{-1}$.
This gave an RMS of 14km s$^{-1}$ for the 58 stars.
Using this uncertainty, the measurements for GL490B are in agreement with
the other measurements, excepting GJ1111.

To check that different spectral features did not give discrepant
values for $\gamma_2$ we could have taken a simple mean 
of the velocities for the five remaining  
objects (four in the case of the higher resolution data) for each
wavelength range.  
However, if we are to exclude GL490B on the grounds it has a poor
determination of its systemic velocity, we must also exclude GJ463.
After removing this star we see
from the RMS (which is, admittedly derived from a small number of
values) that the KI/TiO measurements are in agreement with 
both the high and low resolution NaI velocities.

\subsubsection{Deciding a value and an uncertainty}

Given the above discussion, it seems we can take a very direct
approach and simply take the mean of all the measurements for the 
four good radial velocity standards.
This gives a final measurement of 29$\pm$6 km s$^{-1}$, where the 
uncertainty is simply the standard error.
In principle the errors in the measurements of the velocities of the
standards will add correlations to our 11 measurements, but this is
small compared with our final uncertainty.
We may also have added a little noise by including the high resolution data
(which originate from fewer spectra), but as we shall see later,
the final uncertainty in the gravitational redshift of
the white dwarf is governed not by this error, but by the uncertainty
in $\gamma_1$.

\subsubsection{Comparison with previous work}

One of the most important aims of this work was to decide between the
two contradictory measurements for $\gamma_2$ of $46\pm6$ km s$^{-1}$
\cite{1990MNRAS.246..637F} and $84.9\pm9.9$ \cite{1981ApJ...246..215W}.
Our data clearly favour the \cite{1990MNRAS.246..637F} result, the
difference between the two being at the 2$\sigma$ level.

\subsection{Final dynamical parameters}

We summarise our values for the dynamical parameters as follows.
$K_2$=300 km s$^{-1}$, with an uncertainty which is unclear, but small 
enough to have an insignificant impact on what follows.
$v_2{\rm sin}(i)$=115$\pm$15km s$^{-1}$, where the uncertainty in
the appropriate spectral type to use dominates our error budget.
$\gamma_2$=29$\pm$6 km s$^{-1}$, where the signal-to-noise in our
spectra is probably responsible for the majority of the uncertainty.

\section{PHYSICAL PARAMETERS}
\label{phys}

We can now use our measurements of the dynamical parameters, to derive
the physical parameters of interest.

\subsection{The Mass Ratio}

The mass ratio $M_2/M_1$ is simply the ratio $K_1/K_2$.
\cite{1999ApJ...511..916L} give $K_1=107.1 \pm 2.1$ km s$^{-1}$,
which when combined with our value of $K_2$ yields 0.357$\pm$0.007.

\subsection{The Inclination}
\label{incl}

The eclipse morphology of U Gem gives quite tight constraints on the 
inclination.
The eclipse is thought
to be that of the bright spot and edge of the disc, not the
white dwarf \citep{1971MNRAS.152..219W}, an idea supported by the
delay between spectroscopic phase zero and mid-eclipse
\citep{1981ApJ...246..215W}.
For a $q$ of 0.357, the system would eclipse the white
dwarf if $i>$74 degrees.
For our derived mass ratio we expect the disc to be no larger than
0.66$R_{L1}$ \citep{1977ApJ...216..822P}, a radius which would just be
eclipsed if $i$=62 degrees.
The inclination must, therefore, lie in the range 68$\pm$6 degrees.
Conveniently, the mid-point of this range is close to the 69$\pm$2
degrees given by the most recent of
determinations of the inclination from detailed bright spot
analysis \citep[][and references therein]{2001AcA....51..279S}.

\subsection{The mass of the secondary star}

From Kepler's Law it is easy to show that the mass of the secondary
star is
\begin{equation}
M_2 = {{K_1^3 P}\over{2 \pi G \ {\rm sin}^3(i)}} \left( 1 + {K_2 \over K_1} \right) ^2.
\label{sec_mass}
\end{equation}
Given our values derived above, this gives
\begin{equation}
M_2={0.34\pm0.02M_\odot \over {\rm sin}^3(i)},
\end{equation}
where the uncertainty originates from the uncertainty in $K_1$.
If we wish to derive a value independent of sin$(i)$, then 
taking the range of values of the inclination from above, we find that 
$M_2=0.44\pm0.06$, since we can ignore the uncertainty in $K_1$.
It is worth emphasising that this is an absolute limit, not a 1$\sigma$
uncertainty.

\subsection{The radius of the secondary star}

The radius of the secondary is 
\begin{equation}
R_2 = {{P v{\rm sin}(i)}\over{2 \pi {\rm sin}(i)}} = 0.40 \pm 0.05 R_\odot / {\rm sin}(i).
\end{equation}
In fact the uncertainty in ${\rm sin}(i)$ (4 percent) is smaller than
the uncertainty in our measurement of $v_2{\rm sin}(i)$, so we take the 
mid-point of the range in ${\rm sin}(i)$ to obtain
$R_2 = 0.43 \pm 0.06 R_\odot$.

\subsection{The mass and radius of the primary}

The analogous formula to equation \ref{sec_mass} for the primary is
\begin{equation}
M_1 = {{K_2^3 P}\over{2 \pi G \ {\rm sin}^3(i)}} \left(1 + {K_1 \over K_2} \right) ^2,
\end{equation}
which yields
\begin{equation}
M_1={0.91\pm0.01M_\odot \over {\rm sin}^3(i)}.
\end{equation}
The uncertainty here should be treated with some scepticism as it
includes just the uncertainty in $K_1$, not that from
$K_2$, for reasons discussed in Section \ref{k2_uncer}.
Given our range of inclinations, this leads to a range for $M_1$ of 
1.02 to 1.32$M_{\odot}$.

\begin{figure}
\vspace{70mm}
\includegraphics{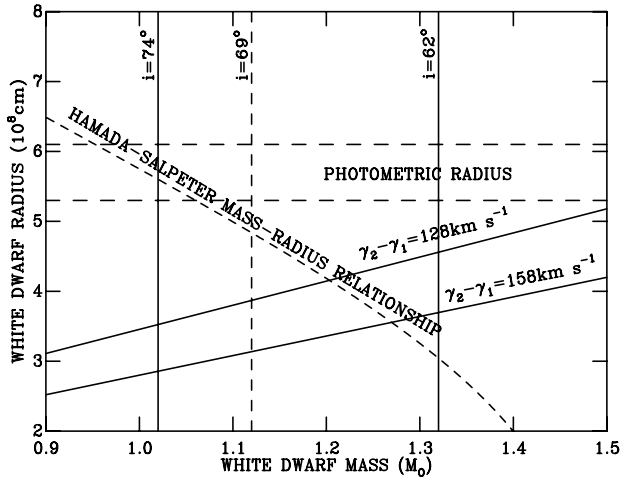}
\caption{
The constraints in the mass-radius plane for the white dwarf.
This is essentially an updated version of Figure 7 of Long \&
Gilliland (1999) and Figure 4 of Long (2000).
The gravitational redshift and inclination constrain the white dwarf
to lie within the box delineated by the solid lines.
The white dwarf mass is primarily limited by the measurement of $K_2$ 
and the inclination limits of 62 to 74 degrees.
(There is, in addition, a weaker dependence on $q$.)
Our measurement of the gravitational redshift to the white dwarf
restricts the white dwarf to lie between the
two lines marked $\gamma_2 - \gamma_1$.
The Hamada-Salpeter mass-radius relationship intersects the box, but
is somewhat above the radius predicted by the inclination of 69
degrees which originates from studies of the bright spot.
The photometric radius (see text) lies between the two horizontal
dotted lines, well above the dynamical limits.
}
\label{wd}
\end{figure}

The radius of the white dwarf can now be determined from the
gravitational redshift ($\gamma_1 - \gamma_2$) as
\begin{equation}
R_1 = {{GM_1}\over{c(\gamma_1-\gamma_2)}},
\end{equation} 
where $c$ is the velocity of light.
\cite{1999ApJ...511..916L} measured $\gamma_1$ as 172$\pm$15km
s$^{-1}$.
If we subtract from this our value for $\gamma_2$ of 29 km s$^{-1}$ we
derive a gravitational redshift for the white dwarf of 143$\pm$15km s$^{-1}$. 
We use this in Figure \ref{wd} to summarise the constraints on $M_1$.
                      
\section{DISCUSSION}
\label{dis}

\subsection{The Secondary Star}
\label{sec_star_disc}

The first question to settle is whether the star lies on the
main-sequence mass radius relationship.
Although small, the uncertainties cover all the mass-radius
relationships given in Figure 2 of \cite{1998MNRAS.301..767S}.
This is still true if we tighten the mass constraint to that given by
using the tighter limits on the inclination given by 
\cite{2001AcA....51..279S}.

\begin{figure}
\vspace{70mm}
\includegraphics{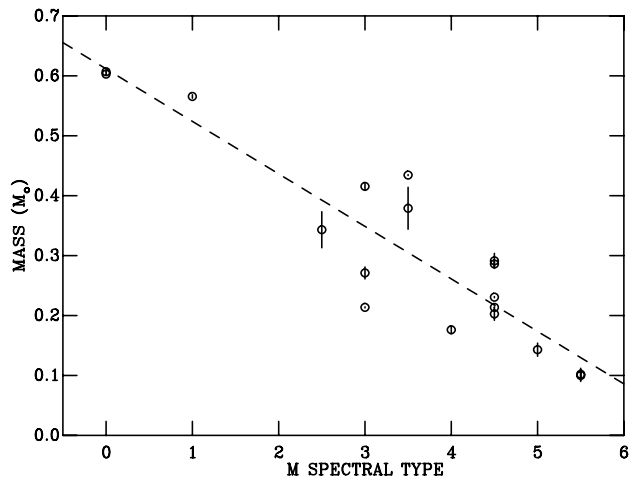}
\caption{
The mass of M-type field stars as a function
of spectral type from Reid et al. (1995) and 
Hawley et al. (1996)
The dotted line is an unweighted linear fit.
}
\label{mass_type}
\end{figure}

Where the problem lies with the secondary star is the spectral type.
In Figure \ref{mass_type} we plot the spectral type of field M-stars given in
\cite{1995AJ....110.1838R} and \cite{1996AJ....112.2799H}
which have masses in \cite{2000A&A...364..217D} against mass.
Given its mass, we would expect the secondary in U Gem to be an M2
star.
Interestingly that is exactly the conclusion reached by 
\cite{2000AJ....120.2649H} on entirely different grounds.
They showed that the absolute magnitude of the secondary star, and 
the optical/IR spectral energy distribution were well matched by
an M2 star with a white dwarf and a faint, power-law accretion continuum.
The problem arises because our best fits of spectral type standards 
to the spectrum of U Gem
imply a spectral type of M4 or M5 \citep[all our spectral types are
also from][]{1995AJ....110.1838R}.
This is perfectly consistent with the results obtained by similar methods by
other workers; M4$^+$ \citep{1990MNRAS.246..637F} and M4.5 
\citep{1981ApJ...246..215W, 1979PASP...91...59S}.

One may be able to solve this inconsistency if the secondary star 
in U Gem has a high metallicity.
Figure \ref{mass_type} shows that there is a spread in spectral type 
at any given mass.
For example, there is an M3.5 star with a mass of 0.415M$_\odot$,
which if it were the analogue of the secondary star in U Gem may
come close to solving the discrepancy.
It is GJ2069A, actually a binary with a mean mass of 0.415M$_\odot$
\citep{1999A&A...341L..63D}, and a spectral type of M3.5
\citep{1995AJ....110.1838R}.
\cite{2000A&A...364..217D} suggest that the late spectral type is due
to high metallicity, with [M/H]=0.5.
One could then explain our cross-correlation results as follows.
If the star is of high metallicity, the lines will be closer to
saturation than one might expect for an M2 star.
As such the line shapes and relative depths of the bandheads will be
closer to that of a later type, solar metallicity star, which has
stronger lines. 
The $\chi^2$ fitting procedure, which attempts to match line shapes
and bandheads (but not the equivalent widths because of the accretion 
continuum) will give a lower $\chi_{\nu}^2$ for the later spectral
type star.
Further evidence that this is the case comes from the fraction of the
light which our fitting procedure gives as originating from the 
secondary star (see Table \ref{results}).
The fraction increases with earlier spectral types, already reaching
100 percent for M3, and presumably would be even higher for an
M2 spectral type standard.
Such behaviour is simply explained if the secondary is an M2 star with 
line strengths enhanced by a high metallicity.

There is one potential problem with the high metallicity explanation, in that
it would also reduce the absolute magnitude.
\cite{2003A&A...398..239R} does indeed find that the absolute $K$-band 
magnitude of GJ2069A is 0.35 mag fainter than stars of comparable mass.
However, such a small change in absolute magnitude corresponds to
about a third of a spectral sub-class, and so would not affect the
conclusions of \cite{2000AJ....120.2649H}, based on the absolute
magnitude of U Gem.

In summary, the observations are consistent with the idea that the
secondary star in U Gem is an M2 dwarf with high metallicity.
In terms of mass, radius and luminosity it is indistinguishable from
similar field M dwarfs.

\subsection{The broader ``cool secondary'' problem}

\begin{figure}
\vspace{70mm}
\includegraphics{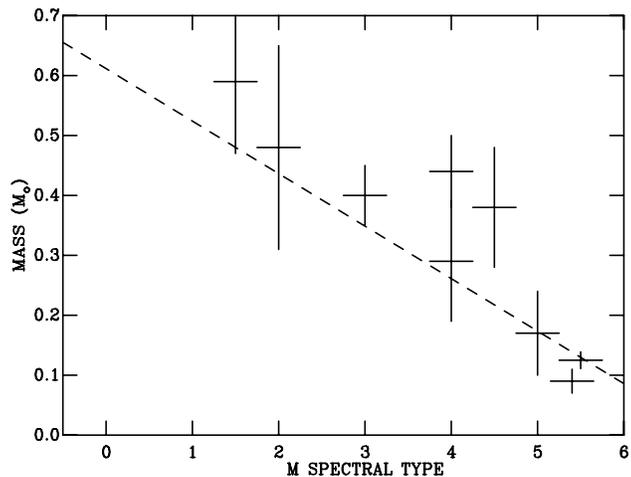}
\caption{
Mass of M-type secondary stars in CVs as a function
of spectral type from Smith \& Dhillon (1998) with
revisions for U Gem from this work and IP Peg from
Beekman et al. (2000).
The dotted line is an unweighted linear fit to the field star
data of Figure \ref{mass_type}.
}
\label{mass_type2}
\end{figure}

There have been suggestions for many years that the secondary stars in
CVs may be too cool for their mass.
If we extrapolate from the observations of U Gem, then the problem is 
not that they are too cool, but simply have a spectral type which is 
later than that for a solar metallicity star of the same effective temperature.
\cite{1998MNRAS.301..767S} apparently showed that such claims were
incorrect, by comparing the mass-spectral-type relation for CV
secondary stars with that for field stars, derived from eclipsing
binaries.
Their argument was that the spread in spectral types of CV secondaries
at a given mass was matched by the spread in field stars.
For the field M-stars, the spread they referred to was driven by the high mass
for the probably-metal-rich GJ2069A.
Conversely, the spread in spectral types for the CV secondary stars
was driven by the high masses of two objects.
First U Gem, whose mass this work revises downwards by 30 percent, and
secondly IP Peg, whose mass we revised downwards by almost a factor
two in \cite{2000MNRAS.318....9B}.
We have plotted therefore, in Figure \ref{mass_type2}, the spectral types
for CVs with well determined secondary star masses from
\cite{1998MNRAS.301..767S} with the revisions for U Gem and IP Peg.

Compared with the data available to \cite{1998MNRAS.301..767S},
there are now many more accurate field star masses, which we have
represented in Figure \ref{mass_type2} by an unweighted linear fit.
The data suggest that, on average, CV secondaries show a tendency 
to be over-massive for their spectral types.
For example, two of the best determinations, IP Peg and U Gem lie at 
later spectral types than any field object of similar mass.
However, this tendency is small (up to about 2 sub-types) and we
agree with \cite{1998MNRAS.301..767S} that it is the spread in both the
field and CV mass/spectral-type relation, rather than any systematic
shift, which precludes the use of mass to spectral-type conversions.

There is current theoretical interest in explaining differences
between the observed spectral type and the spectral type predicted by
the mass of the star \citep[e.g.][]{2001MNRAS.321..544K}.
Our results suggest strongly that any such comparison should be
carried out between the models and the photometric spectral type, not
the spectroscopic one.

\subsection{The White Dwarf}

We can use our data to test whether the white dwarf properties are
similar to those of field white dwarfs.  
From our data we can test whether it follows a similar mass radius
relationship, and whether its luminosity is that expected for its
radius and temperature.

To begin with the mass and radius, we have placed on Figure \ref{wd} 
the Hamada-Salpeter mass-radius relationship as parameterised in 
\cite{1988ApJ...325..266A}.
The range of masses and radii implied by the gravitational redshift in
combination with the estimates of the inclination are consistent with
the Hamada-Salpeter mass-radius relationship.
 To make all three consistent implies that the ``preferred'' inclination
of 69 degrees (see Section \ref{incl}) seems to be a little too high.
However, this conclusion should be tempered with the realisation that
field white dwarfs do not appear to follow this relationship closely.
\cite{1998ApJ...494..759P} show that a large number of white dwarfs 
apparently fall below this relationship, for reasons which are not
clear.
The factor is up to about 20 percent, which would be sufficient to 
solve the discrepancy.

One can also estimate the radius from the FUV spectrum and the
distance, as was discussed by \cite{1999ApJ...511..916L} and
\cite{2000NewAR..44..125L}.
Their results, scaled to reflect the new (and larger) astrometric
distance of 100pc \citep{2004AJ....127..460H} are also plotted in
Figure 8, and indicate a white dwarf radius of $5.7\times10^8$cm,
which we shall refer to as the photometric radius.
The uncertainties in this value are small.
The uncertainty in the parallax is about 5 percent, and hence
contributes 5 percent to the uncertainty in radius.
The overall flux level of the HST spectra is uncertain at the 10
percent level, which translates to a further 5 percent uncertainty in
radius.
This yields an overall uncertainty of 7 percent, or a radius of
$5.7\pm0.4\times10^8$cm.
This figure is inconsistent with our value of the radius derived from
$\gamma_2-\gamma_1$.
This discrepancy is not straightforward to resolve.  
Lower inclinations give a larger white dwarf radius, so if we take the
lowest possible inclination, we still find that $\gamma_2-\gamma_1$
implies a radius of only $4.13\pm0.44\times10^8$cm, which differs from
the photometric radius at the 2.5$\sigma$ level.
Worse still this would place the white dwarf well above the
Hamada-Salpeter radius.
If we wish to make the photometric radius consistent with the
Hamada-Salpeter radius we require $\gamma_2-\gamma_1$=80km s$^{-1}$,
which is far outside our error bars.

One possible solution is to posit another UV component in the 
spectrum \citep{1993ApJ...405..327L}.  
The UV flux in U Gem declines slowly during quiescent intervals.  
However, as was first noted by \cite{1991ApJ...366..569K},
the decline in flux is not consistent with the temperature derived from
modelling the spectra in terms of a uniform temperature white dwarf. 
When modelled as a single temperature white dwarf the temperature of
the white dwarf is hottest and the radius smallest just after
outburst.
\cite{1993ApJ...405..327L, 1995ApJ...454L..39L}
suggested, based on 850-1850\AA\ spectra obtained with the
Hopkins Ultraviolet Telescope, that the discrepancy might be resolved if the
outburst left behind a hot accretion belt on the surface of the white dwarf that
slowly decayed.   
Other possibilities for the second component include
emission caused by heating of the white dwarf due to ongoing accretion and/or
emission from the inner disc.   
The inclusion of a second component does sometimes reduce the
estimated radius of the white dwarf.
However, it is difficult to make such a second component quantitatively
plausible.
For example a white dwarf with a temperature of 30,000 K and a radius
of $4\times10^8$ cm at a distance
of 100 pc would produce only about half of the observed UV flux.  
As a result, to reduce the inferred radius from $5.7 \times 10^8$ to 
$4 \times 10^8$cm requires either that the second component
contributes about the same flux as the first component, or that the
second component causes the temperature of the first component to be
underestimated by order 10,000K, or some combination thereof.  Based on
our experience and the fact that the white dwarf in U Gem is by far the best
studied white dwarf in a dwarf nova, we are sceptical the errors in
the existing spectral
analyses are large enough to easily reconcile the two ways of estimating
the radius of the white dwarf in U Gem.

In summary, the mass and radius of the white dwarf deduced from the
gravitational redshift and binary inclination are consistent with 
those for field stars, but inconsistent with the radius deduced from
the astrometric parallax, temperature and observed UV flux.

\subsection{Determining q using $\bf v_2 sin(i)$}

A technique frequently used in studies of cataclysmic variables is to
measure $q$ using $v_2 {\rm sin}(i)$ and $K_2 $.
This uses the fact that
\begin{equation}
 {{v_2 {\rm sin}(i)}\over{K_2 }}= (1+q){R_2\over a}.
\end{equation}
Since we have a measure of $q$ which is independent of $v_2$sin$(i)$
we can test the validity of this method.
We find that $v_2 {\rm sin}(i) /K_2  = 0.38\pm0.05$.
If we now use the approximation for $R_2/a$ due to
\cite{1983ApJ...268..368E}
we find that $q=0.41\pm0.03$, which is consistent with (though less
accurate than) the value determined using $K_1$, giving
us confidence in parameters determined for this and other systems 
using $v_2 {\rm sin}(i)$. 

\section{Conclusions}
\label{conc}

Our primary conclusions are as follows.
\begin{itemize}
\item{} The mass and radius of the white dwarf in U Gem determined
  from the gravitational redshift and inclination are
  indistinguishable from a field white dwarf.  However, the radius
  deduced from the UV spectrum and astrometric parallax is
  inconsistent with this kinematic determination. Further analysis
  and/or observations are required to understand whether the
  kinematic and spectroscopic information can be reconciled.  This is
  important since U Gem is the proto-typical dwarf nova.
\item{} The secondary star in U Gem is indistinguishable in mass and
  radius from a field M2 dwarf.  However, the spectroscopic spectral
  type is later than might be expected, but this can be explained if
  it is of higher than solar metallicity.
\item{} The M-stars in cataclysmic variables seem, as a group to be
  around 1 spectral sub-type cooler than might be expected from their
  mass or radius.  There is a scatter about the mean mass/spectral
  type relationship of about a sub-type.
\end{itemize}

On our way to reaching these conclusions we have discovered the
following.
\begin{itemize}
\item{} We have tested the determination of the mass ratio using the
  rotational velocity of the secondary star, and shown (at least in
  the case of U Gem) it gives answers consistent with other data.
\item{} However, the uncertainty in the derived rotational velocity is
  dominated by the uncertainty in choosing which spectral type star
  should used for the cross-correlation.
\item{} The uncertainty in the radial velocity semi-amplitude of the
  secondary star is dominated by effects at the few km s$^{-1}$ level
  which are due to the asphericity of the secondary star and
  non-uniformities in the distribution of the line flux over its surface.
\end{itemize}

\section*{Acknowledgements}

The Isaac Newton Telescope is operated on the island of La Palma by the
Isaac Newton Group in the Spanish Observatorio del Roque de los
Muchachos of the Instituto de Astrofisica de Canarias.
We thank Stuart Littlefair for useful discussions, and the referee
Robert Smith for a careful reading and useful suggestions.
Computing was performed on
the Exeter node of the Starlink network, funded by PPARC.

\bibliographystyle{mn2e}
\bibliography{text}

\bsp

\label{lastpage}

\end{document}